\documentclass[final,5p,twocolumn,times,sort&compress]{elsarticle}

\def\preprintdate{September 2018}

\def\tfrac#1#2{{\textstyle{{#1}\over {#2}}}}
\def\dfrac#1#2{{{#1}\over {#2}}}

\def\widehat{\hat}
\def\text#1{{\rm #1}}
\def\sheq{\hskip -6pt &=& \hskip -6pt}
\def\sheqt{\hskip -6.5pt &=& \hskip -6.5pt}
\def\sheqth{\hskip -6.6pt &=& \hskip -6.6pt}

\def\al{\alpha}
\def\be{\beta}
\def\ga{\gamma}
\def\de{\delta}
\def\ep{\epsilon}

\def\et{\eta}

\def\ka{\kappa}
\def\la{\lambda}

\def\ph{\phi}

\def\om{\omega}
\def\Ga{\Gamma}

\def\Om{\Omega}

\def\cl{{\cal L}}

\def\half{{\textstyle{1\over 2}}}

\def\prt{\partial}

\def\lsim{\mathrel{\rlap{\lower4pt\hbox{\hskip1pt$\sim$}}
    \raise1pt\hbox{$<$}}}
\def\gsim{\mathrel{\rlap{\lower4pt\hbox{\hskip1pt$\sim$}}
    \raise1pt\hbox{$>$}}}

\newcommand{\beq}{\begin{equation}}
\newcommand{\eeq}{\end{equation}}
\newcommand{\bea}{\begin{eqnarray}}
\newcommand{\eea}{\end{eqnarray}}
\newcommand{\rf}[1]{(\ref{#1})}
\newcommand{\nn}{\nonumber\\}

\def\mn{{\mu\nu}}

\def\mbf#1{\mathbf{#1}}

\def\ha{(\widehat{k}_a){}}

\def\hc{(\widehat{k}_c){}}
\def\hu{\widehat{u}{}}
\def\hy{\widehat{y}{}}
\def\uu{\overline{u}}
\def\yy{\overline{y}}
\def\F{{F}^{(d)}_1{}}
\def\FF#1{{F}^{(#1)}_1{}}
\def\K{\widetilde{k}^{(d)}{}}

\begin{document}

\begin{frontmatter}

\title{
Riemann-Finsler Geometry and Lorentz-Violating Scalar Fields
}

\author{Benjamin R.\ Edwards and V.\ Alan Kosteleck\'y}

\address{Physics Department, Indiana University, 
Bloomington, Indiana 47405, USA}

\address{}
\address{\rm 
\preprintdate;
published as Phys.\ Lett.\ B {\bf 786}, 319 (2018)
}

\begin{abstract}

The correspondence between Riemann-Finsler geometries 
and effective field theories with spin-independent Lorentz violation 
is explored. 
We obtain the general quadratic action 
for effective scalar field theories in any spacetime dimension
with Lorentz-violating operators of arbitrary mass dimension.
Classical relativistic point-particle lagrangians are derived
that reproduce the momentum-velocity and dispersion relations
of quantum wave packets.
The correspondence to Finsler structures is established,
and some properties of the resulting Riemann-Finsler spaces are investigated.
The results provide support for open conjectures
about Riemann-Finsler geometries 
associated with Lorentz-violating field theories.

\end{abstract}

\end{frontmatter}

\section{Introduction}

A correspondence between 
a large class of Riemann-Finsler geometries
\cite{br,pf}
and realistic effective field theories 
with explicit Lorentz violation
has recently been identified
\cite{ak11}.
The underlying idea is that 
the classical trajectory of a relativistic wave packet 
in the presence of perturbative Lorentz violation
can be mapped via a suitable continuation 
to a geodesic in a Riemann-Finsler space.
The correspondence is of both mathematical and physical interest.
On the mathematics side,
it provides a rich source of examples
of Riemann-Finsler geometries
that are perturbatively close to Riemann geometry.
One example uncovered in this way is a calculable geometry, 
called $b$ space,
that is a natural complement of Randers geometry
\cite{gr}.
The known classification and enumeration 
of Lorentz-violating effects
may also permit a parallel classification
of the corresponding Riemann-Finsler spaces.  
On the physics side,
the correspondence is expected to shed light 
on the poorly understood geometric structure of theories of gravitation
with explicit Lorentz breaking
\cite{ak04}.
Also,
in analogy with the geometric interpretation of Zermelo navigation 
\cite{ez31}
in terms of Randers geometry
\cite{zs03},
the correspondence can be applied to geometric descriptions of physical systems
\cite{fl15}.
Related concepts are explored in various contexts in a broad recent literature 
\cite{ps,cs18,rt18,tbgm,clw,fpp,ilp,sv,nv,em17,dcpm,sma16,rsv,nr15,ms15,js14}.

In nature,
Lorentz and CPT violation could arise 
from an underlying theory combining gravity with quantum physics 
such as strings 
\cite{ksp}.
Observable effects on the behavior 
of known fundamental particles can be inferred
from the comprehensive realistic effective field theory 
for Lorentz violation
incorporating the Standard Model of particle physics 
and General Relativity,
called the Standard-Model Extension (SME)
\cite{dcak,ak04}.
Most of the known fundamental particles have spin,
with only the Higgs boson being a spinless field in the Standard Model.
A nonzero spin complicates the particle trajectory
in part because it involves intrinsically quartic dispersion relations
rather than intrinsically quadratic ones
\cite{kl01}.
However,
even for a particle with nonzero spin,
a subset of Lorentz-violating effects are spin independent
and hence can be handled as though the particle had zero spin.
The combination of relevance and comparative simplicity 
enhances interest in the correspondence between Riemann-Finsler geometries
and the trajectories of particles
experiencing spin-independent Lorentz violation. 

In this work,
we construct the general effective scalar field theories 
in any spacetime dimension 
that contain explicit perturbative spin-independent 
Lorentz-violating operators of arbitrary mass dimension.
The results are used to obtain the general classical lagrangian
describing the propagation of a relativistic spinless point particle
in the presence of Lorentz violation.
The correspondence between the classical lagrangian
and Riemann-Finsler geometries is established,
and some properties of the latter are studied.
Among the results is a set of calculable $y$-global Riemann-Finsler geometries
that are perturbatively close to Riemann geometries.
The properties of these spaces
offer support for some unresolved conjectures about Riemann-Finsler geometries 
associated with Lorentz-violating field theories.

\section{Scalar field theory}

Consider a complex scalar field $\ph(x^\mu)$ of mass $m$
in $n$-dimensional spacetime
with Minkowski metric $\et_\mn$ of negative signature for $n>2$.
The effective quadratic Lagrange density 
describing the propagation of $\ph$
in the presence of arbitrary Lorentz-violating effects
can be written in the form
\bea
\cl(\ph,\ph^\dagger) 
= \prt^\mu\ph^\dagger\prt_\mu\ph - m^2\ph^\dagger\ph 
&&
\nn
&&
\hskip -90pt
- \half\left(i \ph^\dagger\ha^\mu \prt_\mu \ph + {\rm h.c.}\right)
+\prt_\mu \ph^\dagger \hc^{\mn} \prt_\nu \ph ,
\label{L}
\eea
where $\ha^\mu$ and $\hc^\mn$ are operators 
constructed as series of even powers 
of the partial spacetime derivatives $\prt_\al$.
Since Lorentz violation is expected to be small in nature
and perhaps even Planck suppressed,
both $\ha^\mu$ and $\hc^\mn$ can be assumed to introduce only perturbations
to conventional physics.
For some considerations,
it is convenient also to assume that $\ha^\mu$ and $\hc^\mn$
are independent of spacetime position.
This implies translation invariance 
and hence guarantees conservation of energy and momentum,
thereby permitting a focus on Lorentz-violating effects.
The hermiticity of $\cl$ then implies
that $\ha^\mu$ and $\hc^\mn$ 
can be taken as hermitian without loss of generality. 

In the limiting scenario 
in which $\ph$ is a hermitian scalar field,
$\ph^\dagger \equiv \ph$,
the term involving $\ha^\mu$ becomes proportional to 
$i \ph \ha^\mu \prt_\mu \ph + {\rm h.c.}$
However,
all spacetime-constant terms of this type
reduce to total derivatives up to surface terms 
and so in the absence of topological effects
contribute nothing to the classical action.
Note that in the special case of four spacetime dimensions
the term involving $\ha^\mu$ in the theory \rf{L} is CPT odd,
while the one involving $\hc^\mn$ is CPT even.
It therefore follows that
CPT invariance becomes an automatic property
of the propagation of a hermitian scalar field 
in the presence of spacetime-constant Lorentz violation.

The freedom to redefine the canonical variables in a field theory
can imply that certain Lorentz-violating terms in a Lagrange density
are unobservable
\cite{ak04,dcak,kl01,redef,km09,kr10}.
In the present case,
one useful field redefinition takes the form
$\ph \to \ph^\prime = (1 + \widehat{Z})\ph$,
where $\widehat{Z}$ is a Lorentz-violating spacetime-constant operator
formed as a series of powers of derivatives $\prt_\al$.
To preserve the physics of the original theory \rf{L},
which is a perturbation of the free complex scalar field,
the redefinition itself and hence $\widehat{Z}$ must be perturbative. 
Applying the redefinition to the free field theory for $\ph^\prime$
generates perturbative terms proportional to
$\ph^\dagger\widehat{Z}(\prt^\mu\prt_\mu + m^2)\ph+{\rm h.c.}$,
thereby showing that terms of this form
occurring in the Lagrange density \rf{L} 
describe Lorentz-invariant physics
despite their apparent Lorentz-violating form.
It follows that any term in $\ha^\mu$ or $\hc^\mn$
that involves contracted derivatives
can be converted to one with fewer derivatives 
and hence can be absorbed in other terms in the theory \rf{L}.

Under the above assumptions,
the nonderivative pieces of $\ha^\mu$ and $\hc^\mn$ 
can in principle also be removed from the theory \rf{L}.
A spacetime-constant nonderivative component of $\ha^\mu$ is unobservable 
because it can be generated from a conventional free field theory 
using a field redefinition with a nonderivative $\ha^\mu$ of the form
$\ph^\prime = \exp [i\ha_\mu x^\mu/2]\ph$,
which amounts to a position-dependent redefinition of the field phase.
Also,
if $\hc^\mn$ has a nonderivative piece,
it can be absorbed into the metric by a suitable change of coordinates.
However,
in realistic scenarios involving multiple interacting fields 
with distinct nonderivative pieces,
only one combination of pieces can be removed via each of the above methods.
For generality in what follows,
we therefore disregard these options
and instead keep explicitly 
any nonderivative pieces of $\ha^\mu$ and $\hc^\mn$.

The Euler-Lagrange equations of motion for the theory \rf{L} are
\beq
\left(\prt^\mu \prt_\mu + m^2 
+ i\ha^\mu \prt_\mu 
+ \hc^\mn \prt_\mu \prt_\nu \right) \ph =0 .
\label{eqmot}
\eeq
Performing a Fourier transform to momentum space
with the correspondence $p_\mu \leftrightarrow i\prt_\mu$
yields the exact dispersion relation for the theory \rf{L} in the compact form
\beq
p^2-m^2 - \ha^\mu p_\mu + \hc^\mn p_\mu p_\nu =0.
\label{disp} 
\eeq
The operators $\ha^\mu$ and $\hc^\mn$ can conveniently be 
expressed as expansions in even powers of the $n$-momentum $p_\mu$ 
of the form 
\bea
\ha^\mu 
\sheq
\sum_{d\geq n-1}(k_a^{(d)})^{\mu{\al_1}{\al_2}\ldots{\al_{d-n+1}}}
p_{\al_1}p_{\al_2}\ldots p_{\al_{d-n+1}} ,
\nn
\hc^\mn 
\sheq
\sum_{d\geq n}
(k_c^{(d)})^{\mn\al_1\al_2\ldots\al_{d-n}}
p_{\al_1} p_{\al_2} \ldots p_{\al_{d-n}},
\label{acseries}
\eea
where each sum is over either even or odd values of $d$.
The definition of the effective field theory \rf{L} 
for infinite sums over $d$ may be problematic,
so where necessary in what follows we can assume 
the number of Lorentz-violating terms is arbitrary but finite
\cite{km09}.

In Eq.\ \rf{acseries},
the quantities
$(k_a^{(d)})^{\mu\al_1\al_2\ldots\al_{d-n+1}}$
and 
$(k_c^{(d)})^{\mn\al_1\al_2\ldots\al_{d-n}}$
are termed coefficients for Lorentz violation.
They control deviations from conventional propagation
governed by Lorentz-violating operators of mass dimension $d$,
and in physical applications they are the target of experiments
\cite{tables}.
The coefficients have mass dimension $n-d$,
and hermiticity of $\cl$ implies they are real.
The assumption of translation invariance
insures they have constant cartesian components.
Furthermore,
the commutativity of partial derivatives
and the elimination of contracted derivatives via field redefinitions
means that the coefficients can be taken as symmetric and traceless
without loss of generality.
The number $N^{(d)}_n$ of independent components of 
$(k_a^{(d)})^{\mu\al_1\al_2\ldots\al_{d-n+1}}$
or $(k_c^{(d)})^{\mn\al_1\al_2\ldots\al_{d-n}}$
is then found to be
\beq
N^{(d)}_n
=\frac {(2d-n+2)(d-1)!}{(d-n+2)!(n-2)!} .
\eeq
For the special case $n=4$,
this reduces to the standard counting
$N^{(d)}_4=(d-1)^2$
in four spacetime dimensions.

Various limits of the theory \rf{L} can be considered.
For example,
restricting attention to a single nonzero coefficient at a time
can simplify calculations and provide insight.
The coefficients with $d=3$ and $4$ in $n=4$ spacetime
were introduced in Refs.\ \cite{dcak,ak04}
in the context of the Higgs-boson sector of the SME.
The properties of various scalar field theories 
containing these coefficients 
have been widely explored in the literature
\cite{dim34}.
A model with a vector coefficient contributing 
to a $d=6$ term in $n=4$ spacetime
has recently been considered in Ref.\ \cite{npr},
but other scenarios with $n\neq 4$ or $d>n$
appear unexplored to date.
 
Another unexplored limit of potential interest 
allows only coefficients with timelike indices to be nonzero.
This model is spatially isotropic in the defining inertial frame,
although spatial anisotropies arise in most other frames.
Note that spatial isotropy could in principle also be achieved
by tracing over spatial components,
but the requirement that all coefficients are traceless
in any pair of spacetime indices
implies that a pair of traced spatial indices 
can be replaced with a pair of timelike indices 
without loss of generality.
At each value of $d$,
the spatially isotropic limit therefore allows only one coefficient,
denoted by $k^{(d)}$.
The dispersion relation for this model 
then takes the comparatively simple form
\beq
E^2 - |\mbf p|^2 - m^2 + \sum_d (-1)^{d-n} k^{(d)} E^{d-n+2} = 0 ,
\eeq
where $E$ is the energy of the particle of spatial momentum $\mbf p$,
and where the sum is over all values of $d\geq n-1$.

\section{Classical kinematics}

The behavior of a wave packet obeying the equations of motion \rf{eqmot}
is controlled by the dispersion relation \rf{disp},
which describes the effects of Lorentz violation
on the energies of plane waves of different momenta.
The dispersion relation can alternatively be interpreted
as the energy-momentum relation
for an analogue classical point particle.
The motion of this analogue particle
is determined by a lagrangian $L$,
which in turn can be related to Finsler geometry.
The construction of $L$ for a given dispersion relation
is therefore of definite interest.
Although obtaining an explicit result for $L$ can be challenging, 
a formal procedure to achieve this has been given in Ref.\ \cite{kr10}.
Here,
we extend this procedure to $n$ spacetime dimensions
and develop an iterative method to calculate $L$ explicitly.
For definiteness in what follows,
we assume $m\neq 0$.
Also, where appropriate and convenient we write
$(k^{(d)})^{\mu\nu\al_1\al_2\ldots\al_{d-n}}$
for either $-(k_a^{(d)})^{\mu\nu\al_1\al_2\ldots\al_{d-n}}$
or $(k_c^{(d)})^{\mn\al_1\al_2\ldots\al_{d-n}}$,
which simplifies expressions 
that contain coefficients with both $a$ and $c$ subscripts 
or that are valid for either alone.

The motion of the analogue particle follows 
a worldline in $n$ dimensions. 
The worldline can be parametrized by $\la$ 
and specified by the $n$ equations $x^\mu = x^\mu(\la)$,
and the $n$-velocity $u^\mu$ of the particle 
is then given by $u^\mu=dx^\mu/d\la$.
In the general case
$L$ depends on both the position and the velocity of the particle,
but the assumption of translation invariance of $k$
implies that $L=L(u,k)$ is independent of the position
and that the canonical $n$-momentum 
$p_\mu = -\prt L/\prt u^\mu$
is conserved.
Invariance of the action under reparametrizations of $\la$
requires that $L$ be homogeneous of degree 1 in $u^\mu$.
Applying Euler's theorem then reveals that $L$
can be written implicitly as $L=-u^\mu p_\mu$.
The relation between the $n$-momentum and the $n$-velocity
is fixed by matching the spatial velocity of the analogue particle 
to the group velocity of the wave packet in the field theory,
$-{u^j}/{u^0} = {\prt p_0}/{\prt p_j}$.
The challenge of constructing an explicit expression for $L(u,k)$
such that the Euler-Lagrange equations 
reproduce the dispersion relation \rf{disp}
then reduces to solving simultaneously 
the $n-1$ matching equations and the dispersion relation 
to obtain $p_\mu$ in terms of $u^\mu$.

For simple cases,
an analytical solution for $L=L(u,k)$ can be found.
Consider,
for example,
the field theory with only one particular nonvanishing coefficient,
$(k^{(n)}_c)^\mn \ne 0$.
The dispersion relation can then be written as
$p_\mu \Om^\mn p_\nu=m^2$,
where $\Om^\mn=\et^\mn+(k^{(n)}_c)^\mn$.
Taking the derivative of the dispersion relation
with respect to $p_j$ yields 
$(u^0\Om^{j\nu} - u^j\Om^{0\nu})p_\nu =0$.
Multiplying by $p_j$ and some manipulation of the result
provides an implicit expression for the $n$-velocity,
$u^\mu =-L^{(n)}\Om^\mn p_\nu/m^2$.
Since the coefficients $(k^{(n)}_c)^\mn$ are assumed perturbative,
the inverse of the matrix $\Om$ exists.
Left multiplication of the implicit expression for $u^\mu$
with $u^\al(\Om^{-1})_{\al\mu}$
then yields an expression for $(L^{(n)})^2$.
Identifying the physical root 
by requiring that the usual result is recovered
in the limit $(k^{(n)}_c)^\mn \rightarrow 0$
reveals that
\beq
L^{(n)}(u,k^{(n)}_c)=-m\sqrt{u^\mu(\Om^{-1})_\mn u^\nu}.
\label{lnuk}
\eeq
For the special case $n=4$,
this matches the result in Ref.\ \cite{kr10}.
Notice that smoothness of $L^{(n)}$ fails 
for any $n$-velocity 
for which $u^\mu(\Om^{-1})_\mn u^\nu=0$.
This reflects the deformation of the light cone
introduced by the coefficients $(k^{(n)}_c)^\mn \ne 0$,
and it parallels the failure of smoothness
of the standard free-particle lagrangian $L(u)=-m\sqrt{u^\mu\et_\mn u^\nu}$
when $u^\mu\et_\mn u^\nu=0$.

For field theories having coefficients with $d\geq n+1$,
the dispersion relation can still be written in the form
$p_\mu \Om^\mn p_\nu=m^2$,
but with $\Om^\mn = \Om^\mn(p)$ now a function of the $n$-momentum. 
Following the above procedure then leads to a higher-order polynomial in $L$
for which explicit solution is typically impossible.
Nonetheless,
the intermediate steps provide useful implicit expressions.
Incorporating arbitrary $k$ coefficients for $d\geq n+1$,
we find an implicit expression for the $n$-velocity to be
\bea
u^\mu
\sheqt
-L p_\nu
\big[ \et^\mn
\nn
&&
\hskip 10pt
+ \half \sum_d (d-n+2)
p_{\al_1}\ldots p_{\al_{d-n}}
(k^{(d)})^{\al_1\ldots\al_{d-n}\mn} 
\big]
\nn
&&
\hskip -10pt
\times 
\big[m^2+\half \sum_d (d-n)
p_{\al_1}\ldots p_{\al_{d-n+2}}
(k^{(d)})^{\al_1\ldots\al_{d-n+2}}
\big]^{-1} ,
\label{uprel}
\eea
where the sums are over all values of $d\geq n+1$.
Contraction with $u_\mu$ 
yields a quadratic polynomial for $L$,
the solution of which gives the implicit expression 
for the lagrangian as
\bea
L
\sheq
- \big[ m^2 \uu^2
\nn
&&
+\tfrac{1}{16}\big(\sum_d (d-n+2)
u_{\al_1} p_{\al_2} \ldots p_{\al_{d-n+2}}
(k^{(d)})^{\al_1\ldots\al_{d-n+2}}\big)^2 
\nn
&&
+ \half\sum_d (d-n) \uu^2
p_{\al_1}\ldots p_{\al_{d-n+2}}
(k^{(d)})^{\al_1\ldots\al_{d-n+2}}
\big]^\half 
\nn
&&
\hskip -6pt 
+ \tfrac{1}{4} \sum_d (d-n+2)
u_{\al_1} p_{\al_2} \ldots p_{\al_{d-n+2}}
(k^{(d)})^{\al_1\ldots\al_{d-n+2}} ,
\label{Lroot}
\eea
where $\uu\equiv \sqrt{u^\mu\et_\mn u^\nu}$.

Direct manipulation of the results \rf{uprel} and \rf{Lroot}
to extract $L(u,k)$ is infeasible in many cases.
However,
we can develop an iterative method 
that generates the solution as a series
in powers of the coefficients for Lorentz violation.
The idea is to expand both implicit expressions
for $L(u,p,k)$ and $u^\mu$ as power series in $k$
and then to perform successive substitutions 
to derive an expression for $L(u,k)$
valid at the chosen order in $k$. 
For simplicity,
we illustrate the method 
in the special case of a model with only one nonzero coefficient
$(k^{(d)})^{\al_1\ldots\al_{d-n+2}}$
of mass dimension $n-d$,
denoting the resulting lagrangian by $L^{(d)}$.
However,
the results presented below can be generalized
to more complicated scenarios as desired. 

The first step is to expand the implicit lagrangian \rf{Lroot} 
in powers of $k$.
For the chosen model,
we find
\bea
L^{(d)}
\sheq
\tfrac{1}{4}(d-n+2)
u_{\al_1} p_{\al_2} \ldots p_{\al_{d-n+2}}
(k^{(d)})^{\al_1\ldots\al_{d-n+2}} 
\nn
&&
\hskip -16pt 
+m \uu \sum_{s=0}^q \sum_q 
(-1)^q a_{qs} 
\nn
&&
\hskip 10pt
\times
[(d-n+2)
u_{\al_1} p_{\al_2} \ldots p_{\al_{d-n+2}}
(k^{(d)})^{\al_1\ldots\al_{d-n+2}}]^{2s} \nn
&&
\hskip 10pt
\times
[(d-n)(p_{\al_1}\ldots p_{\al_{d-n+2}}
(k^{(d)})^{\al_1\ldots\al_{d-n+2}})]^{q-s} ,
\label{keyseries}
\eea
where 
\beq
a_{qs} = 
\frac{(2q)! }
{m^{2q}\uu^{2s}(2q-1)8^{q+s}q!s!(q-s)!} .
\label{aqs}
\eeq
Given $L^{(d)}=L^{(d)}(u, p, k^{(d)})$ expressed as Eq.\ \rf{keyseries},
the iteration then proceeds as follows.
The zeroth-order lagrangian 
$L^{(d)}_0\equiv L^{(d)}(u,p,0)=-m\uu$
is defined as the limit of vanishing $k^{(d)}$.
The corresponding zeroth-order momentum is 
$(p_0)_\mu \equiv -\prt L^{(d)}_0/ \prt u^\mu = m u_\mu/\uu$.
The $q$th-order lagrangian is then defined
by inserting the $(q-1)$th-order momentum into Eq.\ \rf{keyseries}, 
$L^{(d)}_q = L^{(d)}(u,p_{q-1}(u),k^{(d)})$,
keeping only terms up to the $q$th power of $k^{(d)}$.
The $q$th-order momentum is obtained in the canonical way by differentiation,
$(p_q)_\nu=-\prt L^{(d)}_q / \prt u^\nu$.

This iteration method shows that $L^{(d)}(u,k^{(d)})$ 
can be determined to any order in $k^{(d)}$
and that the explicit relationship
between the $n$-momentum and the $n$-velocity
is obtained at each step.
Smoothness of the lagrangian $L^{(d)}(u,k^{(d)})$ 
outside the usual slit $S_0 \equiv \{ u^\mu | \uu=0 \}$
is then insured at any order in Lorentz violation.
Note that the derivation of the result \rf{keyseries}
involves expanding the radical in the implicit expression \rf{Lroot},
which implies the allowed values of $k^{(d)}$ are constrained.
A first-order form of the constraint is obtained by
demanding that the magnitude of the ratio of the summands in the radical
is bounded above by unity 
and inserting the zeroth-order momentum $(p_0)_\mu$,
giving
$|(k^{(d)})^{\al_1\ldots\al_{d-n+2}}
\hu_{\al_1} \ldots \hu_{\al_{d-n+2}}|<2/(d-n)$
for $d>n$.
Potential convergence issues arising in the limit 
of an infinite sum over $d$
are tied to the corresponding definition of the effective field theory 
in that limit
and hence are moot in the present context.

As an illustration in the context of the chosen model,
we present here the results of a calculation using this iterative method
applied to third order in the coefficient for Lorentz violation.
The third-order lagrangian is found to be 
\bea
L^{(d)}_3 
\sheq
L^{(d)}_0
\big[
1 - \half \K
- \tfrac 18 (d-n+1)^2(\K)^2
\nn
&& 
\hskip -6pt 
+ \tfrac 18 (d-n+2)^2 
~\K_\al ~\K^\al
- \tfrac 1{16} (d-n+1)^4(\K)^3
\nn
&&
\hskip -6pt 
+ \tfrac 1{16} (d-n+1)(2d-2n+1)~\K ~\K_\al ~\K^\al
\nn
&&
\hskip -6pt 
- \tfrac 1{16} (d-n+1)(d-n+2)^2~\K_\al ~\K^{\al\be}~\K_\be 
\big],
\label{thirdL}
\eea
where we have introduced the dimensionless quantities 
\beq
\K_{\al_1\ldots\al_l} = 
m^{n-d} (k^{(d)})_{\al_1\ldots\al_l\al_{l+1}\ldots\al_{d-n+2}}
\hu^{\al_{l+1}}\ldots \hu^{\al_{d-n+2}} 
\eeq
with $\hu^\al \equiv u^\al/\uu$.
This expression is indeed smooth away from $\uu=0$,
as expected.
Note that although the derivation assumes $d\geq n+1$,
the results also hold for $d=n-1$ and $d=n$.
For the former case,
the expression \rf{thirdL} directly matches the analytical result.
For the latter case,
all the quantities $a_{qs}$ 
given in Eq.\ \rf{aqs} vanish except when $s=q$,
leaving the expected third-order approximation
to the exact result \rf{lnuk}.

In the context of the general field theory 
for the propagation of Dirac fermions in $n=4$ spacetime dimensions
in the presence of arbitrary Lorentz violation
\cite{km13},
Reis and Schreck used an ansatz-based technique
to obtain the corresponding classical lagrangian for the analogue particle 
at leading order in coefficients for Lorentz violation
\cite{rs18}.
The resulting effects of spin-independent Lorentz violation 
can be expected to match those of the theory \rf{L}
because the latter contains all possible spin-independent effects
for a propagating particle.
Indeed,
we can confirm that a match exists to the first-order part 
of the expression \rf{thirdL} with $n=4$,
via the correspondences 
$\hat{a}^{(d)}_\star 
\leftrightarrow 
\half(k_a^{(d)})_{\al_1\ldots \al_{d-2}}u^{\al_1}\ldots u^{\al_{d-2}}$
and 
$\hat{c}_\star^{(d)} 
\leftrightarrow 
\half(k_c^{(d)})_{\al_1\ldots \al_{d-2}}u^{\al_1}\ldots u^{\al_{d-2}}$.
Substituting these correspondences into the full expression \rf{thirdL}
is therefore expected to generate the third-order lagrangian 
describing spin-independent Lorentz-violating effects
on the propagation of a Dirac particle. 
Similar results can be anticipated
for spin-independent Lorentz-violating effects
on photon
\cite{km09}
and neutrino
\cite{km12}
propagation as well.

\section{Finsler geometry}

A classical reparametrization-invariant point-particle lagrangian 
is a smooth real-valued function on the slit tangent bundle
that is 1-homogeneous in the velocity
and that yields the equation of motion via a variational principle. 
Its features have parallels with those of a Finsler structure
underlying a Riemann-Finsler geometry,
with key differences being the signature of the metric
and the requirement of positivity.
These differences could conceivably be obviated 
via a suitable definition of Lorentz-Finsler geometry,
producing a relationship to Riemann-Finsler geometry
analogous to that between Lorentz and Riemann geometry.

To date,
no completely satisfactory and widely accepted definition
of Lorentz-Finsler geometry exists.
Various approaches have been suggested including,
for example, 
those in 
Refs.\ \cite{jb70,gsa85,ma94,bf00,ak11,pw11,lph12,js18}.
However,
relaxing the positivity requirement in a consistent way
while including all natural physical examples of point-particle lagrangians 
remains an elusive goal.
For instance,
a sophisticated recent effort
is the causality-based construction of Javaloyes and S\'anchez
\cite{js18},
which succeeds in incorporating special cases of the $a$ and $b$ lagrangians 
derived from effective field theory with Lorentz violation
\cite{ak11}
and also exposes sharply the challenge of finding a definition
that includes other related physical examples.

The results of Sec.\ 3 above 
play two primary roles in the context of Finsler geometry
\cite{ak11}.
First,
we can promote the Minkowski metric $\et_{\mn}$ 
to a spacetime metric $r_{\mn}(x)$ 
and allow position dependence of the coefficients.
This procedure generates all the classical lagrangians controlling
dominant effects on the spin-independent propagation of a particle 
in a general spacetime background perturbed by arbitrary Lorentz violation.
It thereby substantially increases the known physical examples
offering potential guidance in the search 
for a suitable definition of Lorentz-Finsler geometry.

Second,
independently of the definition of Lorentz-Finsler geometry,
we can focus instead on the issue 
of generating a Finsler structure for a Riemann-Finsler geometry 
from a classical point-particle lagrangian with Lorentz violation.
The existence of this relationship is of direct interest in is own right,
particularly since a subset of the mathematical properties
derived for the lagrangian formulation
can be expected to transfer to the Riemann-Finsler geometry.
In the present context,
the results obtained in Sec.\ 3
can be used to generate all Riemann-Finsler geometries 
associated with spin-independent Lorentz violation
that are perturbations of conventional Riemann geometry.
The classification and enumeration of these lagrangians
is therefore expected to establish
a corresponding classification of Riemann-Finsler spaces
that are perturbatively related to a Riemann space.

Several methods can be countenanced
to establish the desired relationship.
The most direct procedure amounts to defining a suitable analytic continuation 
of the spacetime coordinates and derivatives, 
the coefficients for Lorentz violation,
and the lagrangian,
thereby yielding directly
a Finsler structure for a Riemann-Finsler geometry.
This method has some features in common with a Wick rotation
in quantum field theory,
and we adopt it in what follows.
Other possible approaches could include
converting the original quantum field theory
to its euclidean counterpart via analytic continuation
and then performing an analysis in parallel
with that in Sec.\ 3 above,
or implementing a projection or truncation 
of the spacetime to the purely spatial subspace 
and suitably adapting the classical lagrangian.
Investigation of these alternative options
and of their uniqueness and potential equivalence
would be of interest but lies outside our present scope. 

\bigskip
Starting with the results presented in Sec.\ 3,
the continuation is implemented via the mappings
$u^\mu \rightarrow i^N y^j$,
$p_\mu \rightarrow (-i)^N p_j$,
$(k^{(d)})^{\mu\ldots} \rightarrow i^N (k^{(d)})^{j\ldots}$,
and 
$L \rightarrow -F = -y\cdot p$.
For convenience and to match conventions in the Riemann-Finsler literature,
we also impose $m \to 1$.
In these expressions,
the $n$ spacetime dimensions $x^\mu$ labeled with Greek indices 
are replaced with $n$ spatial dimensions $x^j$ labeled with Latin indices.
Also,
$N$ is a generic symbol representing the number of spacelike indices 
present in a quantity prior to its continuation.
For instance,
the timelike component of $u^\mu$ acquires a factor of 1,
while each spacelike component acquires a factor of $i$.
The resulting expressions can naturally be rewritten 
using the euclidean metric.
We can then promote this metric to a Riemann metric $r_{jk}(x)$
and allow spacetime dependence of the coefficients,
in parallel with the procedure discussed above for the spacetime case. 
As an example,
this produces the map
$\sqrt{u^\mu \et_{\mn} u^\nu} 
\rightarrow \sqrt{y^j r_{jk}(x) y^k}$.

\bigskip
When implemented on the broad set of classical lagrangians
associated to effective field theories with Lorentz violation,
the above procedure yields Finsler structures for Riemann-Finsler geometries 
that are perturbatively related to a Riemann space
\cite{ak11}.
Note that the original theory \rf{L}
is defined using components $(k^{(d)})^{\mu\ldots}$
of tensors in $\otimes TM$
and hence generates a Finsler structure
in terms of tensor components $(k^{(d)})^{j\ldots}$.
Starting instead with cotensor or mixed-tensor components 
generates a family of distinct Finsler structures
related by factors of $r_{jk}$.
However,
the $y$ dependence is unaffected,
so the results obtained below for $(k^{(d)})^{j\ldots}$
can be directly transcripted to any other desired member of the family.

\bigskip
Using this technique,
the results in Sec.\ 3 
for the velocity-momentum relation \rf{uprel} 
and the classical lagrangian \rf{Lroot}
become implicit expressions for the Finsler structure
of a Riemann-Finsler geometry,
\bea
y^j
\sheqth
\frac{F p_k
\left[ r^{jk}
+ \half \sum_d (d-n+2) p_{l_1}\ldots p_{l_{d-n}}
(k^{(d)})^{l_1\ldots l_{d-n}jk} 
\right]}
{1+\half \sum_d (d-n)p_{l_1}\ldots p_{l_{d-n+2}}
(k^{(d)})^{l_1\ldots l_{d-n+2}}} 
\nn
\label{impyeq}
\eea
and 
\bea
F
\sheq
\big[ \yy^2
\nn
&&
+\tfrac{1}{16}\big(\sum_d (d-n+2)
y_{l_1} p_{l_2} \ldots p_{l_{d-n+2}}
(k^{(d)})^{l_1\ldots l_{d-n+2}}\big)^2 
\nn
&&
+\half \yy^2 \sum_d (d-n)
p_{l_1}\ldots p_{l_{d-n+2}}(k^{(d)})^{l_1\ldots l_{d-n+2}}
\big]^\half 
\nn
&&
\hskip -6pt 
-\tfrac{1}{4} \sum_d (d-n+2)
y_{l_1} p_{l_2} \ldots p_{l_{d-n+2}}
(k^{(d)})^{l_1\ldots l_{d-n+2}},
\label{Fseries}
\eea
where $\yy=\sqrt{y^j r_{jk} y^k}$. 
The result \rf{Fseries} is smooth 
on the slit bundle $TM$$\setminus$$S$,
where $S=S_0+S_1$ contains the usual slit $S_0$ containing $y^j=0$
but is extended to include other roots 
of the expression in the radical above. 
This geometry is therefore generically $y$ local,
although for certain restrictions
the geometry may be resolvable along $S_1$.

Direct solution of the above implicit results
to yield an explicit expression for $F$
is typically impractical.
To extract an explicit result,
we can instead parallel the iteration procedure described in Sec.\ 3
and thereby generate the $q$th-order Finsler structure $F_q$.
In this context,
it is natural to define
\beq
\K_{j_1\ldots j_l} = 
(k^{(d)})_{j_1\ldots j_l j_{l+1}\ldots j_{d-n+2}}
\hy^{j_{l+1}}\ldots \hy^{j_{d-n+2}} 
\label{def}
\eeq
with $\hy^j \equiv y^j/\yy$,
as the iteration introduces these combinations.
Note that the indices on
$(k^{(d)}){}^{j_1\ldots j_l j_{l+1}\ldots j_{d-n+2}}$
are lowered using the Riemann metric $r_{jk}$.
Also,
the number of indices on $\K_{j_1\ldots j_l}$
reveals the number of contractions with $\hy^j$.
For example, 
$\K$ denotes contraction of all indices. 
As before,
the iteration procedure involves an expansion in powers of the coefficients
of the radical in Eq.\ \rf{Fseries}.
Requiring the magnitude of the ratio of the summands in the radical
to be bounded above by unity at first order
and inserting the unperturbed momentum
$p_j=\hy_j=y^j/\yy$
yields the constraint $|\K| < 2/(d-n)$.

At first iteration order and keeping coefficients of arbitrary $d$,
the iteration produces the compact expression 
\beq
F_1=\yy-\half \yy\sum_d \K.
\label{firstorder}
\eeq
This Finsler structure is smooth 
on the usual slit bundle $TM$$\setminus$$S_0$,
so the corresponding geometry is $y$ global.
Indeed,
the same is true for $F_q$ at any finite $q$
because the process generates a series of terms 
in powers of $\K_{j_1\ldots j_l}$,
and the latter is smooth away from $y^j=0$.
Note that $F_q$ for any given $q$ can be viewed
either as generating an approximation to the full geometry
implied by Eqs.\ \rf{impyeq} and \rf{Fseries}
or as an independent Finsler structure 
yielding a $y$ global geometry of interest in its own right.
Note also that $F_1$ is reversible,
$F_1(y)=F_1(-y)$,
iff $(k_a^{(d)})^{j_1\ldots}$ vanishes.
This property holds at any iteration order as well.
Reversibility of a Riemann-Finsler geometry
corresponds to CPT invariance in effective field theories with $n=4$.
Imposing it eliminates half of the allowed values 
of $d$ in Eq.\ \rf{firstorder}.

At higher iteration orders, 
the mixing of coefficients of different $d$ makes $F_q$ unwieldy.
Also,
the Finsler metric $g_{jk}=(F^2)_{y^j y^k}/2$
involves derivatives of the square of $F_q$,
which introduces further mixing and yields burdensome expressions.
To gain insight through direct calculations,
it is therefore useful to consider 
the special case with only one nonzero coefficient.
For example, 
the explicit form of the third-order Finsler structure $F^{(d)}_3$ 
can be found immediately by continuation from Eq.\ \rf{thirdL}.
However, 
for our purposes below it suffices to limit attention 
to the first-order Finsler structure
\beq
\F=\yy-\half \yy ~\K.
\label{f1d}
\eeq
In the context of the above discussion,
the derivation of this expression assumes $d>n$.
However,
for the case $d=n$ we can understand 
$\FF n$ as implementing a linearized shift of a conventional Riemann metric,
while for the case $d=n-2$ we see that
$\FF {n-2}$ is merely the usual Riemann geometry with a scaled mass.
Also,
for the case $d=n-1$,
inspection reveals that $\FF {n-1}$ 
is the standard Randers structure
built with the 1-form $(k^{(n-1)}_a)_j y^j/2$.
We can therefore extend the interpretation 
of Eqs.\ \rf{firstorder} and \rf{f1d}
to $d \geq n-2$ when desired.
Note that from this perspective
the Finsler structure $\F$ with $d>n$
can be viewed as a natural generalization
of the Randers structure,
in which the 1-form is replaced by a symmetric $(d-n+2)$-form. 
In a similar vein,
$\F$ can be viewed as a generalization
of the Finsler structure for a geometry with an $(\al,\be)$ metric,
in which $\al\equiv\yy$ and the 1-form $\be$ is generalized
to a symmetric $(d-n+2)$-form. 

To verify that $\F$ is indeed a Finsler structure,
certain conditions must be met
\cite{bcs}.
One is positive homogeneity in $y^j$,
which is evident by inspection.
Another is smoothness on the usual slit bundle $TM$$\setminus$$S_0$,
which holds as already noted above.
A third is nonnegativity,
which is achieved when $1-\half\K>0$.
This condition is automatically satisfied when $|\K|<2/(d-n)$,
which is the constraint obtained above
from expanding the radical in Eq.\ \rf{Fseries}.

Another condition is positivity of the Finsler metric.
Imposing this can be expected to translate
into an additional constraint on $\K_{j_1\ldots j_l}$
in terms of $d$ and $n$. 
Here,
we derive this constraint explicitly 
at leading order in 
$\K_{j_1\ldots j_l}$.
At this order,
we find the Finsler metric $g_{jk}^{(d)}$ is given by
\bea
g_{jk}^{(d)}
\sheq
r_{jk} [1 + \half (d-n) \K]
-\half(d-n+1)(d-n+2)\K_{jk}
\nn
&&
\hskip -15pt 
+\half(d-n)(d-n+2)
(\K_j \hy_k + \K_k \hy_j - \K \hy_j \hy_k).
\label{metric}
\eea
As expected,
for $d=n$ this result represents a simple scaling of $r_{jk}$,
while for $d=n-1$ it reduces to the linearized Randers metric.

An argument for positivity of the metric \rf{metric}
can be made in terms of the positivity of its determinant $\det g^{(d)}$,
so we first consider the latter.
It is convenient to define
$\ka \equiv \max\{|k^{(d)}_{j_1\ldots j_{d-n+2}}|\}$.
We then find
$|\K| = |(k^{(d)})_{j_1\ldots j_{d-n+2}}\hy^{j_1}\ldots \hy^{j_{d-n+2}}|
\le n^{d-n+2}\ka$.
Similarly, 
$|\K^j{}_j| \le |r^j{}_j| ~n^{d-n}\ka=n^{d-n+1}\ka$.
Writing $g^{(d)}_{jk}=r_{jk}+h_{jk}$
implies $\det g = (1+ h^j{}_j) \det r$ at first order,
where the trace is with respect to $(r^{-1})^{jk}$.
The triangle inequality then yields the relation
\bea
&&
\hskip -20pt
|(d-n+1)(d-n+2)\K^j{}_j-(d-n)(d+2)\K| 
\nn
&&
\hskip -25pt
< (d-n+1)(d-n+2)|\K^j{}_j| + (d-n)(d+2)|\K| 
\nn
&&
\hskip -25pt
< (d-n+1)(d-n+2)n^{d-n+1}\ka + (d-n)(d+2)n^{d-n+2}\ka.
\nn
\eea
It follows that $\det g^{(d)}>0$ at linear order when
\beq
\ka < \frac{2}{ [(d-n+1)(d-n+2)+(d-n)(d+2)n]n^{d-n+1}}.
\eeq
For $d>n$,
the smallest value of $d$ is $d=n+1$,
which gives $\ka < 2/n^2(n^2+3n+6)$.
For example,
if the Finsler structure is derived 
from a field theory in $(3+1)$ spacetime dimensions,
then $n=4$ and the smallest value $d=5$
imposes $\ka < 1/272$.
More generally,
this shows that for any case with $d>n$
sufficiently small coefficients can be found
that ensure positivity of $\det g^{(d)}$
at linear order.
A standard argument 
\cite{bcs}
then suffices to show positivity of the metric at linear order.
Introducing $F_{1\ep}^{(d)}=\yy-\half \yy\ep\K$,
it follows from the above argument that $\det g_\ep^{(d)} >0$,
and so $g_{\ep jk}^{(d)}$ has no vanishing eigenvalues.
Since $g_{\ep jk}^{(d)} \to r_{jk}$ with positive eigenvalues 
when $\ep \to 0$,
the eigenvalues must stay positive as $\ep \to 1$,
and so $g_{jk}^{(d)}$ must be positive definite at linear order.
For sufficiently small $\ka$,
we expect positivity to hold 
at higher orders in $\K_{j_1\ldots j_l}$ as well,
but a formal proof of this remains open at present.

\section{Some properties of $k$ spaces}

Next,
to gain insight about the various Riemann-Finsler spaces
governed by $(k^{(d)})^{j_1\ldots j_{d-n+2}}$,
we perform some explicit calculations for the Finsler structure \rf{f1d}.
The expressions for key properties below are derived 
at first order in $\K_{j_1\ldots j_l}$.

Consider first the Hilbert form $\om \equiv F_{y^j} dx^j$
for a given Finsler structure $F$.
This is a section of the pullback bundle $\pi^* T^* M$ 
defined globally on the usual slit bundle $TM$$\setminus$$S_0$
\cite{bcs}.
The components $p_j= F_{y^j}$ are the Riemann-Finsler analogues 
of the components of the $n$-momentum per mass
in the corresponding classical lagrangian.
A short calculation for the Finsler structure $\F$ reveals
\beq
p^{(d)}_j = [1+\half(d-n+1)\K] ~\hy_j - \half(d-n+2)\K_j .
\eeq
In the Riemann limit with $\K\to 0$,
$p^{(d)}_j \to \hy_j$ is aligned with the velocity.
The presence of nonzero $(k^{(d)})^{j_1\ldots j_{d-n+2}}$
scales this result in a direction-dependent way
and shifts it by a direction-dependent covector,
so that $p^{(d)}_j$ and $\hy_j$ generically become linearly independent.

For $d>n$,
we can show explicitly 
that none of the Riemann-Finsler $k$ spaces with Finsler structure $\F$
are Riemann geometries.
The noneuclidean aspects of a Finsler structure $F$
interpreted as a Minkowski norm on any tangent space $T_xM$
are captured by the Cartan torsion
$C_{jkl}\equiv (g_{jk})_{y^l}/2$,
which according to Deicke's theorem 
\cite{deicke}
vanishes only for Riemann geometries. 
We find that the first-order Cartan torsion is 
\bea
C^{(d)}_{jkl} 
\sheq
\frac{1}{4\yy}(d-n)(d-n+2)
\nn
&&
\times \sum_{(jkl)}\big[ 
\left( 
\tfrac{1}{3}(d-n+4) \hy_j \hy_k \hy_l - r_{jk} \hy_l 
\right) 
\K 
\nn
&&
\hskip 30pt
-\tfrac{1}{3}(d-n+1)\K_{jkl}
+ (d-n+1)\K_{jk}\hy_l 
\nn
&&
\hskip 30pt
+ [r_{kl}-(d-n+2)\hy_k \hy_l ] \K_j 
\big] .
\label{cartan}
\eea
This Cartan torsion vanishes for $d=n$ and $d=n-2$,
in agreement with our earlier identification
of $\FF n$ and $\FF {n-2}$ as Finsler structures for Riemann geometries.
Inspection reveals that the Cartan torsion also vanishes for $n=1$,
as is appropriate for a Riemann curve.
However,
the mean Cartan torsion is nonzero for other values of $n$ and $d$,
indicating that in those cases the Finsler structures $\F$ 
cannot correspond to Riemann geometries.
In the reversible scenario with $(k_a^{(d)})^{j_1\ldots}=0$
on a compact surface,
this implies the Finsler metric has nonconstant flag curvature
\cite{az88,rb07}.
In the nonreversible scenario for $\F$,
a Finsler metric with constant positive flag curvature 
may exist and would be interesting to display 
\cite{bfimz}.

The Cartan torsion \rf{cartan} is nonvanishing for $d=n-1$.
The corresponding space is identified in Sec.\ 4 as a Randers geometry.
According to the Matsumoto-H\=oj\=o theorem 
\cite{mh},
Randers spaces are distinguished by a nonvanishing Cartan torsion
together with a vanishing Matsumoto torsion
$M_{jkl}=C_{jkl}- \sum_{(jkl)}I_jh_{kl}/(n+1)$,
where $I_l = g^{jk}C_{jkl}$ is the mean Cartan torsion
and $h_{jk}=F (p_k)_{y^j}$ is the angular metric.
For the Finsler structure $\F$,
the mean Cartan torsion at first order is 
\bea
I^{(d)}_j
\sheq
\frac{1}{4\yy}(d-n)(d-n+2) \big[
(d+2)( \K_j - \K \hy_j ) 
\nn
&&
\hskip 40pt 
-(d-n+1)( \K^k{}_{jk} - \K^k{}_k \hy_j )
\big].
\eea
Calculation then yields the first-order Matsumoto torsion as
\bea
M^{(d)}_{jkl}
\sheq
\frac{(d-n)(d-n+1)(d-n+2)}{4(n+1)\yy}
\nn
&&
\times \sum_{(jkl)}
\big[
\tfrac{1}{3}(n-2)\K \hy_j \hy_k \hy_l 
+ r_{jk} (\K \hy_l - \K_l)
\nn
&&
\hskip 20pt 
- n \K_l \hy_j \hy_k
- (r_{kl} - \hy_k \hy_l) 
(\K^m{}_m \hy_j - \K^m{}_{mj})
\nn
&&
\hskip 20pt 
+ (n+1)(\K_{jk} \hy_l - \tfrac{1}{3} \K_{jkl}) 
\big].
\eea
For the Randers value $d=n-1$,
this expression indeed vanishes.
The Matsumoto torsion $M^{(d)}_{jkl}$ 
also vanishes for the Riemann values $d=n$ and $d=n-2$,
as expected.
However,
it is nonvanishing for $d>n$,
which establishes that none of the corresponding spaces are Randers geometries.
Note also that this result holds for $n=2$,
which implies the $k$ spaces with $d>2$ must be distinct from $b$ space
because the latter reduces for $n=2$ to Randers geometry
\cite{ak11}.
This distinction is consistent 
with the different nature of the $k$-space and $b$-space coefficients
as bases for representations of the rotation group O($n$)
and might be anticipated because $b$ space is related 
to spin-dependent Lorentz violation,
unlike the $k$ spaces.
A similar argument suggests the $k$ spaces differ
from other Riemann-Finsler geometries
related to spin-dependent Lorentz violation,
including the various $H$ spaces considered in Ref.\ \cite{krt12}.

Another approach to Riemann-Finsler geometry is through geodesic sprays
\cite{zs-sprays}.
For any choice of speed or diffeomorphism gauge,
a Riemann-Finsler geodesic is a solution of the equation
\beq
F\frac{d}{d \la}\big( \frac{y^j}{F} \big) + G^j=0 ,
\eeq
where 
$G^j = g^{jk} (\prt_l g_{km} + \prt_m g_{kl} - \prt_k g_{lm}) y^l y^m/2$
are the spray coefficients.
Denoting the Christoffel symbol for $r_{jk}$
by $\widetilde{\ga}^j{}_{kl} 
= \half r^{jm}(\prt_k r_{lm}+\prt_l r_{km}-\prt_m r_{kl})$
and the covariant derivative with respect to $r_{jk}$
by $\widetilde{D}_j$,
some calculation reveals that the first-order spray coefficients 
for the Finsler structure $\F$ can be expressed as
\bea
\frac{1}{\yy^2}G^{(d)j}
\sheq
\widetilde{\ga}^j{}_{\bullet\bullet}
+ \half \widetilde{D}^j \K
+ \half(d-n) \hy^j \widetilde{D}_\bullet \K 
\nn
&&
\hskip 20pt 
- \half(d-n+2)r^{jk}\widetilde{D}_\bullet\K_k .
\label{kspray}
\eea
Here,
a bullet $\bullet$ indicates contraction of a lower index $j$ with $\hy^j$,
and all contractions with $\hy^j$ are understood 
to be taken outside any derivatives.
Note that the spray coefficients
are homogeneous of degree two in $y^j$.

The expression \rf{kspray} reveals the noteworthy result
that if the coefficients are $r$-parallel,
$\widetilde{D}_k ({k}^{(d)}){}^{j_1\ldots j_{d-n+2}}=0$,
then the first-order spray coefficients reduce to Riemann ones
and hence the presence of $r$-parallel 
$({k}^{(d)}){}^{j_1\ldots j_{d-n+2}}$
leaves unaffected the geodesic curves.
It turns out that the analogous result also holds 
for $ab$ and $face$ spaces
\cite{ak11}.
Taken together,
these results support the conjecture that any $r$-parallel coefficient 
leaves Riemann geodesics unaffected.
Local conditions along a geodesic appear sufficiently uniform 
in a geometry with $r$-parallel coefficients 
that nonzero $({k}^{(d)}){}^{j_1\ldots j_{d-n+2}}$
cannot be observed. 
At the level of the effective field theory,
these results lend weight to the open possibility
of removing $r$-parallel coefficients
using field redefinitions and coordinate choices
similar to those already used to remove unphysical coefficients
in certain limits of the SME
\cite{ak04,dcak,kl01,redef,km09,kr10}.

The spray coefficients can be used to derive
many useful quantities in Riemann-Finsler geometry
\cite{mm86aim,bcs}.
One is the nonlinear connection,
which can be defined as $N^j{}_k\equiv (G^j)_{y^k}/2$.
Using this definition and the homogeneity properties
of the spray coefficients,
we can write $G^j=y^k N^j{}_k$.
For the Finsler structure $\F$,
a calculation reveals that 
the first-order nonlinear connection $N^{(d)j}{}_k$
takes the form
\bea
\dfrac{1}{\yy} N^{(d)j}{}_k
\sheq
\widetilde{\ga}^j{}_{\bullet k}
+ \tfrac{1}{4}(d-n)
(\de^j{}_k \widetilde{D}_\bullet \K
+ \hy^j \widetilde{D}_k \K)
\nn
&& 
\hskip -6pt 
+\tfrac{1}{4}(d-n)(d-n+2)\hy^j
(\widetilde{D}_\bullet \K_k-\hy_k\widetilde{D}_\bullet \K ) 
\nn
&&
\hskip -6pt 
-\tfrac{1}{4}(d-n+2)r^{jl}\big[
\widetilde{D}_k \K_l - \widetilde{D}_l \K_k 
\nn
&& 
\hskip 50pt 
-(d-n) ( 
\hy_k\widetilde{D}_\bullet \K_l 
- \hat{y}_k\widetilde{D}_l \K
)
\nn
&& 
\hskip 50pt 
+(d-n+1)\widetilde{D}_\bullet \K_{kl}
\big] .
\eea
Note that this expression reduces to its Riemann equivalent
for $r$-parallel coefficients 
$({k}^{(d)}){}^{j_1\ldots j_{d-n+2}}$.

Various connections for Riemann-Finsler geometry 
can be derived from the nonlinear connection. 
One is the Berwald connection
${}^B\Ga^j{}_{kl}=(N^j{}_l)_{y^k}$.
Its explicit form for the Finsler structure $\F$ 
is somewhat cumbersome,
so we omit it here.
However,
the expressions given above imply that 
${}^B\Ga^{(d)j}{}_{kl}=\widetilde{\ga}^j{}_{kl}$
for $r$-parallel coefficients 
$({k}^{(d)}){}^{j_1\ldots j_{d-n+2}}$.
It follows that the Berwald h-v curvature defined as
${}^BP_j{}^k{}_{ml} = -F({}^B\Ga^k{}_{ml})_{y^j}$
vanishes in this case.
The $r$-parallel $k$ spaces of this type are therefore Berwald spaces.
This result adds further support to the open conjecture
that any SME-based Riemann-Finsler space is a Berwald space
iff it has $r$-parallel coefficients
\cite{ak11},
which was previously proved for Randers space
\cite{mm,hi,ssay,sk}.

Another quantity of importance is the Chern connection
\cite{sc48},
which is defined as
$\Ga^j{}_{kl}\equiv
\half g^{jm}
(\de g_{mk} / \de x^l
+ \de g_{ml} / \de x^k
- \de g_{kl} / \de x^m)$,
where $\de / \de x^j = \prt_{x^j} -N^k{}_j \prt_{y^k}$
is the dual to $dx^k$.
For the Finsler structure $\F$,
some calculation reveals the first-order expression
\bea
\Ga^{(d)j}{}_{kl}
\sheq
\widetilde{\ga}^j{}_{kl} 
\nn
&&
\hskip -6pt 
+ \tfrac{1}{4}r^{jm}\big[
(d-n)[r_{km}-(d-n+2)\hat{y}_k\hat{y}_m] \widetilde{D}_l \K 
\nn
&&
\hskip 20pt
+ (d-n)(d-n+2) [\hat{y}_l \widetilde{D}_k\K_m + (l\leftrightarrow m)]
\nn
&&
\hskip 20pt
-(d-n+1)(d-n+2) \widetilde{D}_k \K_{lm} 
\nn
&&
\hskip 20pt
+ (k \leftrightarrow l)-(k\leftrightarrow m)
\big].
\eea
This reduces to the usual Levi-Civita connection
for $r$-parallel coefficients.

The Chern connection
can be used to calculate the Chern curvatures.
Consider,
for example,
the Chern h-v curvature
defined as $P_k{}^j{}_{ml}\equiv -(\Ga^j{}_{ml})_{y^k}$.
After some calculation with the Finsler structure $\F$,
we find that the components 
$P^{(d)}_{jklm} \equiv g_{ks}P^{(d)}_j{}^s{}_{lm}$
are given at first order by
\bea
P^{(d)}_{jklm}
\sheq
\dfrac{1}{4\yy}(d-n)(d-n+2)
\nn
&&
\hskip -10pt 
\times
\sum_{(jkl)} 
\big[
( r_{kl} - \tfrac{1}{3}(d-n+4) \hat{y}_k \hat{y}_l ) 
\hat{y}_j \widetilde{D}_m \K 
\nn
&&
\hskip 15pt 
-[r_{kl}-(d-n+2) \hat{y}_k \hat{y}_l] \widetilde{D}_m\K_j 
\nn
&&
\hskip 15pt 
- (d-n+1) \hat{y}_j \widetilde{D}_m \K_{kl} 
\nn
&&
\hskip 15pt 
+\tfrac{1}{3}(d-n+1) \widetilde{D}_m \K_{jkl} 
\big]
\nn
&&
+(l\leftrightarrow m) -(k\leftrightarrow l). 
\eea
This result reveals that the Chern h-v curvature vanishes
when the coefficients 
$({k}^{(d)}){}^{j_1\ldots j_{d-n+2}}$
are $r$ parallel.

As a final remark,
we note that other widely used connections
can also be calculated for the $k$ spaces
using the above results.
For example,
the Cartan connection can be defined as 
$\Ga^j{}_{ml}dx^l + C^j{}_{ml}\de y^l$,
where we can take the Cartan tensor to be given by 
$C^j{}_{ml} = r^{js}C_{sml}$ at first order
and where $\de y^j = dy^j + N^j{}_k dx^k$
is the dual to $\prt_{y^j}$.
Another example is the Hashiguchi connection,
which can be defined as 
${}^B\Ga^j{}_{ml}dx^l + C^j{}_{ml} \de y^l$.
However,
these connections fail to reduce to the Riemann result
even for $r$-parallel coefficients 
$({k}^{(d)}){}^{j_1\ldots j_{d-n+2}}$
due to the nonvanishing Cartan tensor.

\section*{Acknowledgments}

We thank J.F.\ Davis, M.A.\ Javaloyes, C.\ Judge, P.\ Kirk, 
Z.\ Shen, N.\ Voicu, and W.\ Ziller
for useful discussions.
This work was supported in part
by the United States Department of Energy 
under grant number {DE}-SC0010120
and by the Indiana University Center for Spacetime Symmetries.

\end{document}